\documentclass[12pt,preprint]{aastex}

\shorttitle{A MERLIN Observation of PSR B1951+32 \& Plerion}
\shortauthors{Golden et al.}

\begin{document}


\title{A MERLIN Observation of PSR B1951+32 and its associated Plerion}


\author{A. Golden\altaffilmark{1}, 
S. Bourke\altaffilmark{1}, G. Clyne\altaffilmark{1}, R.F. Butler\altaffilmark{1}, 
A. Shearer\altaffilmark{1},T.W.B. Muxlow\altaffilmark{2},W.F. Brisken\altaffilmark{3}}

\altaffiltext{1}{Computational Astrophysics Laboratory, I.T. Building,
National University of Ireland, Galway; agolden,stephen,ger,ray,shearer@it.nuigalway.ie}
\altaffiltext{2}{Jodrell Bank Observatory, University of Manchester, Cheshire
SK11 9DL, UK; twbm@jb.man.ac.uk}
\altaffiltext{3}{National Radio Astronomy Observatory, P.O. Box O, Socorro, NM 87801; wbrisken@nrao.edu}


\begin{abstract}
In an investigative 16 hour L band observation using the
MERLIN radio interferometric array, we have resolved both the 
pulsar PSR B1951+32 and structure within the flat spectral radio 
continuum region, believed to be the synchrotron nebula associated 
with the interaction of the pulsar and its `host' supernova remnant
CTB 80. The extended structure we see, significant at
$\sim$ 4.5 $\sigma$, is of dimensions
2.5" $\times$ 0.75", and suggests a sharp bow shaped arc
of shocked emission, which is correlated with similar
structure observed in lower resolution radio
maps and X-ray images. Using this MERLIN data as a new
astrometric reference for other multiwavelength data we  
can place the pulsar at one edge of the HST reported optical
synchrotron knot, ruling out previous suggested optical
counterparts, and allowing an elementary analysis of the
optical synchrotron emission which appears to trail the 
pulsar. The latter is possibly a consequence of
pulsar wind replenishment, and we suggest
that the knot is a result of magnetohydrodynamic (MHD) instabilities.
These being so, it suggests a dynamical nature to the optical knot,
which will require high resolution optical observations to confirm. \\
\end{abstract}



\keywords{ISM: individual (CTB 80) --- ISM: supernova remnants --- pulsars: individual (PSR B1951+32) --- stars:neutron 
--- radiation mechanisms: non-thermal}


\section{Introduction}

Recent subarcsecond Hubble Space Telescope (HST) \& Chandra 
observations of the inner Crab
Nebula have yielded evidence of dynamic
activity in close proximity to the pulsar
with various shocks and `wisps' evolving 
in terms of  position, morphology and luminosity over timescales 
of days to weeks, implying local velocities of order 0.7c
(Hester et al. 1996; Weisskopf et al. 2000).
Observations in the radio using the Very Large Array (VLA) show
evidence for similar time-varying morphological 
changes within the nebula but at scales of order 1" (Bietenholz et al. 2001).

Such multiwavelength 
observations challenge our understanding of these \\
pulsar/plerion/supernova remnant (SNR) associations.
Considering the Crab Nebula, the estimated current
rate of particle injection derived via nebular
X-ray luminosity differs markedly from the 
historical average rate determined from the radio
emitting particles (Arons, 1998; Atoyan, 1999).
The nebula's emission may be modelled as 
a function of the particle injection spectrum
related to the spin down energy of the pulsar,
yet the exact processes which link the thermalising
pulsar wind to the observed synchrotron emission
are not clear (Reynolds \& Chevalier, 1984), neither 
is our understanding of the particle acceleration 
processes within the pulsar wind in the first place 
(Gallant \& Arons, 1994; Lou, 1993).  

Another `young' pulsar/SNR/X-ray plerion association that 
provides a working laboratory with which to test our
understanding of such interactions following on from the
Crab is that of the young radio, X-ray
and $\gamma$-ray pulsar PSR B1951+32 associated
with the CTB 80 SNR.
Early radio and
X-ray observations noted a central plerion/spectrally flat
region to the SW of the SNR,  within which PSR B1951+32 was 
detected, located within a concentration of nebular
emission towards one edge of the flat spectral region
(e.g. Strom et al. 1984). The association
is valid based on this clear interaction between
pulsar and remnant, with similar pulsar canonical
age (107 kyr)  and dynamically derived SNR age 
(9.6 $\times$ 10$^{4}$d$_{2.5kpc}$). 

Strom (1987) performed a detailed survey
of this central flat region with the VLA at various
frequencies and baselines, all yielding maps with
resolutions $\sim$ 1", and
recent surveys using the WSRT at 6 cm 
(Strom \& Stappers, 2000) have
indicated considerable complexity within the spectrally flat
core. Strom  was the first to comment on the `hot spot',
located to the SW of the pulsar, suggesting that it
was likely a consequence of the pulsar wind interacting
with the remnant and creating a `wisp' like structure
similar to that observed in close proximity to the
Crab pulsar. 

In contrast to the young Crab, it appears that the older
PSR B1951+32 has caught up with its expanding remnant,
resulting in the observed complex multiwavelength emission.
As such, it represents an extremely interesting stage
in the life cycle of a pulsar, when the neutron star
penetrates and interacts with the remnant/swept up ISM region.
Precisely what radiation gets emitted, and where, will define the
constraints to any subsequent modelling efforts.

Moon et al. (2004) have recently reported an optical/X-ray
analysis of the system using archival HST and Chandra data,
in addition to new ground based optical and IR observations.
The Chandra data clearly shows a cometary pulsar wind nebula
which appears to be confined by a bow shock produced by the
high velocity motion of the pulsar, which corresponds to Strom's
previously defined radio `hot spot'. Optical/IR photometry have also indicated the presence
of a synchrotron 'knot', as originally reported by Hester (2000) 
embedded within the cometary head observed by Chandra.
Previous
attempts to isolate putative optical counterparts to PSR B1951+32
have been concentrated around this structure, which is believed
to be closely related to the pulsar - thus far two candidates
have been proposed which are within the VLA error ellipse (Butler
et al. 2002). Moon et al. (2004), based on their analysis, argue that
only one is likely, and even then it is generally agreed that 
the `point source' involved may well be a background star or
non-uniformity in the knot itself.

The key to resolving many of the outstanding aspects of the 
the system's multiwavelength geometry is a rigorous astrometric
reference frame. 
In this short letter we report a 16 hour L band observation 
with the MERLIN radio interferometer.
Both objects were resolved at a resolution
of 150 mas, and despite the relatively low signal to noise of the extended
emission associated with the shock front, we have been able
to re-examine existing multiwavelength data using the MERLIN
data as the astrometric reference.

\section{Observations \& Data Analysis}
On January 19th 2002, a 16 hour L-band MERLIN\footnote{MERLIN, a UK National 
Facility  operated by the University of Manchester at Jodrell Bank  
Observatory on behalf of PPARC}
observation (without the Lovell Telescope) was performed
on the central spectrally `flat' region centred on PSR B1951+32. 
The phase calibrator used was 1951+355, whose position
is tied into the ICRF, with a positional accuracy 
of $<$ 0.5 mas.
The correlator was configured to operate at 1658 MHz
with a total bandwidth of 16 MHz on 32 channels. 
The data was processed and analysed within the AIPS
evironment. 
Initial maps
showed contamination from a bright off-axis radio source.
This was identified as MG3 J195211+3248 ($\alpha ~ \sim$ 19:52:15.7811, 
$\delta ~ \sim$ 32:49:36.299), a symmetric
double radio galaxy, with a measured flux of $\sim$ 80 mJy.
Its contribution was subsequently CLEANed out.

\subsection{Imaging the Pulsar and Environs}
The pulsar is clearly evident in figure 1, and its position as obtained
during this observation with MERLIN is determined to be
$\alpha$ $\sim$ 19:52:58.204 ($\pm$0.002$\arcsec$), $\delta$ $\sim$ +32:52:40.531 ($\pm$0.025$\arcsec$) (2452294 JD). 
PSR B1951+32 was detected as a point source in this observation, 
and there was no evidence for scatter-broadening. 
There is also a significant ($\sim$ 4.5 $\sigma$) detection of
a structure within the anticipated `hot spot' some 3" SW
of the pulsar, of approximate
dimensions 2.5" $\times$ 0.75", with the pulsar's motion approximately
bisecting the observed emission. 

Despite being relatively weak in
terms of signal to noise, we note that this `bead-like' arc of emission 
corresponds with lower-resolution data obtained with the VLA\footnote{The National Radio 
Astronomy Observatory is a facility of the National Science Foundation operated under 
cooperative agreement by Associated Universities, Inc.} 
We estimate that an excess flux density
of $\sim$ 2.84 mJy along the narrow (1-2 beams in width) linear
feature on top of the larger ($\sim$ 1") structure.
This is similar to our determination of $\sim$ 2.59 mJy of
excess flux relative to the shell from the complementary VLA data.
Whilst the errors of these measurements is likely to be larger
than their difference, this suggests that the shock structure is
not over resolved in the MERLIN observations, and that the
actual emission region is spatially limited.
Thus, whilst we are confident that MERLIN has resolved fine-structure
within the `hot spot' observed previously, although given the low signal to
noise ratio, one must be cautious to avoid over-analysing the MERLIN data.

\subsection{The Optical Synchrotron Knot}
The optical knot, first identified by Hester (2000) in the original
HST F547M observations, can now be placed in its correct astrometric
context given this MERLIN dataset. To do this, we must astrometrically calibrate the
HST image to a significantly better degree than what is 
provided via the GSC. This is done by referencing the HST
data to the 2MASS Point Source Catalog. The limiting astrometric accuracy of 
the latter is $\sim$ 100-150 mas (1 sigma) relative to the Hipparcos/Tycho 
coordinate system, which realizes the ICRS for optical data.

The image was astrometrically calibrated to the ICRS via the
following fitting process. 40 stars common to both the WFPC2
and the 2MASS Point Source Catalog were identified, the former
extracted as pixel coordinates using DAOPHOT and were then corrected for 
the known geometric distortions. These 
pixel coordinates are each internally accurate to $<<$100 mas, 
since the stars have a PSF FWHM of $\sim$125 mas.

The initial matching was based on the pipeline World Coordinate 
System (WCS) of the HST image. The first fit showed that this 
WCS was approx. 500 mas systematically in error. A second matching and 
fitting iteration was performed, 2nd-order CCMAP fits used a 2-sigma rejection 
cutoff to remove input stars of lower astrometric fidelity. 
The final iteration resulted in 9 of the 40 input stars being 
rejected and 31 stars surviving to define the fit for a new WCS 
solution, with a final fit rms of 105 mas in RA and 101 mas in Dec. 
This refined WCS was then written into the header of the HST 
image.

Comparison with the MERLIN astrometry is thus performed purely 
by reference to this WCS. Therefore, we stress that the 
radio-optical comparison is in no way dependent on any dubious 
procedure such as trying to tie together one or more features 
which might appear to be in common to the HST and MERLIN images. 
Their respective calibrations to the ICRS were performed via 
completely independent and robust means, and are anchored
within the WCS schema. 

In figure 2 we overlay the 2MASS
corrected HST image on our L band MERLIN map, using the
WCS header information in both cases. Hester's knot is
apparent in the field centre. Morphologically, the knot resembles a teardrop-like 
structure, of dimensions 0.8" $\times$ 1.3". We also
show the position of the radio pulsar and
the shock front. Astrometric integration of the
MERLIN data yields an overall accuracy in registering the optical to the 
radio frame of $\sim$ 105 mas. 
Within the combined astrometric errors as defined by
the WCS solutions, this alignment makes a powerful case
for the pulsar and Hester's nebula to be associated.
The pulsar is located at the lower south-west extent of the optical
knot region. Knowing the approximate direction of the pulsar's motion, this
suggests that the knot's observed optical emission is a consequence of shocked
pulsar wind material and/or shocked nebular material during the transit
of PSR B1951+32. 

Ram pressure balance between 
relativistic pulsar winds and the ambient medium may be represented
as $\rho_{a} \nu^{2}_{\rm psr} \sim \dot{E}/4 \pi \Omega c r^{2}_{s}$,
with $\rho_{a}$ the ambient medium density, $\dot{E}$ the pulsar's
spin-down energy, the $4 \pi \Omega$ term the solid angle of the 
pulsar wind flow and $r_{s}$ the stagnation radius - $c$ being the
speed of light. Typically in these treatments, the ram balance
implies equipartition of particle and magnetic field pressure at the
shock front, and this allows one to express the local magnetic field
as $B_{eq}(\mu G) \sim 50(n_{H}/{\rm cm^{3}})^{1/2} (v_{psr}/100 {\rm km s^{-1}}))$, 
or equivalently $B_{eq}(\mu G) \sim 200 \Omega^{-1/2} (\dot{E}/10^{36} 
{\rm erg s^{-1})}^{1/2}(r_{s}/0.01 {\rm pc}^{-1})$
with $n_{H}$ the hydrogen number density. 

The recent VLA observations
of Migliazzo et al. (2002) have determined $v_{psr}$ = 240 km s$^{-1}$.
Chandra data imply $r_{s}$ = 0.03 pc (or 3" at 2 kpc) 
and $\dot{E}$ = 3.7 $\times$ 10$^{36}$ ergs s$^{-1}$.
Assuming $\Omega$ $<$ 1, Moon et al. (2004) argue that $B_{eq}$ $>$ 100 $\mu$G.
These authors indicate that this result correlates with earlier Hester \& Kulkarni (1989)
results of a pre-shock $n_{H}$ of 50 cm$^{-3}$ and $B_{eq}$ $\sim$ 600 $\mu$G.
Moon et al. (2004) then argue that if one assumes $B$ $>$ 100 $\mu$G in the X-ray
nebula, then the synchrotron cooling time of these X-ray photons is
$t_{\rm sync} \sim 40 E^{-1/2}_{\rm keV} (B/100 \mu G)^{-3/2}$ or $<$ 40 yrs.
Given the nebula's spatial extent along the pulsar's axis of motion 
of $\sim$ 20", and PSR B1951+32's proper motion of
$\sim$ 25 mas/yr, this points towards the nebulae being constantly replenished
by the pulsar wind, certainly in X-rays.

Optically, the knot's length 
of $\approx$ 1.3" and the pulsar's proper motion suggest
a lifetime of $\sim$ 50 years. 
Assuming the knot's emission is a consequence of `cooling' synchrotron
particles, one can use this crossing time as the cooling time, and the energy of these
optical photons to constrain the previous equation to yield a value for
$B_{eq}$, which comes to $\sim$ 600 $\mu$G, 
not inconsistent with the previous studies.

However, this analysis suggests that the
observed optical emission is entirely explicable as synchrotron emission
from the pulsar wind particles left behind after passage of the pulsar -
without the requirement of particle replenishment by the same pulsar
wind. Moon et al. 2004 used particle replenishment 
to justify their analysis of the Chandra data.  
For replenishment to be valid (Kaspi et al. 2001), the 
freshly shocked wind particles must be continuously fed `backwards'
with a velocity that is high enough given their cooling times.
Given the estimated cooling times and dimensions of the optical knot,
an X-ray knot associated with it would have a particle cooling time
of $\sim$ 0.3 to 9 years, with corresponding particle flow velocities
of $\sim$ 185 - 260 to km s$^{-1}$ (optical) and 1.5 $\times$ 10$^{3}$ to 5 $\times$
10$^{4}$ km s$^{-1}$ (X-ray).

An alternative interpretation is that the observed optical 
synchrotron knot is a consequence of quasi-stationary 
shock structures in the pulsar wind outflow behind the 
pulsar (Lou 1998). At 
various spin latitudes, reverse fast MHD shocks can appear 
quasistationary when their propagation speeds relative to 
the pulsar wind are comparable to the relativistic outflow. 
The most likely source of disturbances which explain why 
these wisps and knots appear where they are observed are 
slightly inhomogeneous wind streams 
emanating from the rapidly spinning pulsar. A slower wind 
stream will eventually be caught up by a faster wind stream 
to trigger forward and reverse fast MHD shock waves in the 
distant pulsar wind. 

Enhanced synchrotron emission is expected from these knots 
due to the pitch-angle scattering of MHD shock-energised 
relativistic particles. There are various mechanisms available 
to energise particles to high energies from plasma turbulence, 
from both the forward and reverse fast relativistic MHD shocks 
themselves as well as magnetic reconnection in the pulsar wind.
Collectively, they strongly suggest the likelihood of a dynamic
component to this optical knot.

Considering that PSR B1951+32 possesses
an estimated transverse velocity of $\sim$ 240 km s$^{-1}$, a counter knot may be 
`smothered' by the bow shock formed  by the pulsar's passage 
towards the SW of the remnant. At the bow shock 
itself injected particles from the pulsar wind are advected behind the pulsar 
while some of the particles diffuse across the bow shock and into the shell 
of the SNR thereby rejuvenating the shells emission (van der Swaluw et al. 2002, 
Shull et. al 1989), which we see to the south-west of this remnant.  
Therefore the MHD wind streams may be advected away before they have an 
opportunity to interact, and therefore do not form a knot in the scenario 
proposed by Lou (1998).
Taking into account recent simulations (Buccaniti 2002, Gaensler et al. 2004), 
the knot seems to lie in the subsonic region that exists behind the pulsar and it 
is surrounded by the supersonic region created by the back flowing material from 
the bow-shock. At the interface between these two regions there may exist shearing 
forces which may explain the asymmetry of the optical knot which is evident in
figure 2.

\subsection{Implications for Proposed Optical Counterparts to PSR B1951+32}

There are strong grounds for the detection of an optical counterpart
to PSR B1951+32, based on empirically derived relationships between other known
optical pulsars (Shearer \& Golden, 2001), and also on the fact the
pulsar is a known X-ray and putative $\gamma$-ray pulsar. A previous
analysis of this same HST data by Butler et al. (2002) suggested two
potential optical counterparts which were coincident with the best
reported VLA astrometry at that time - one a clear point source to the
SW of the knot, the other embedded within the knot towards its upper
NW edge. Moon et al. (2004) state that the former counterpart does not
satisfy more rigorous astrometry, and propose that the latter is
the only remaining candidate, albeit with the reasonable caveat that
this latter source may be a background star or a transient localised
emission region within the knot - a point also made by Butler et al. (2002). 
Our MERLIN astrometry rules out
any association between it and the pulsar. However, it is reasonable
to consider the possibility that the proposed counterpart is (was?) a non-uniformity within
the synchrotron nebula. This would in turn support the idea that the
optical structure we observe is a direct consequence of the pulsar
wind outflow, and again argues the case for the knot to be a 
dynamically evolving structure.

\section{Conclusions}

We report the first high resolution radio observation of the inner
PSR B1951+32 plerion using MERLIN at L band.We have resolved both the pulsar
and apparent fine structure within the `hot spot' identified at lower
resolution and believed to be a consequence of the pulsar wind interacting
with swept up ISM/SNR material.

We have used our MERLIN data to register the astrometrically corrected 
archival HST observations of the field. Combined, these data indicate 
that the previously identified optical `knot' of synchrotron emission 
extends behind the pulsar, along a line that bisects the shock front emission.
The dimensions of the optical knot and the VLA determined 
proper motion argue for a synchrotron cooling time that is consistent with particle
replenishment from the pulsar wind. 

The formation of the knot can also be attributed
to the mechanisms outlined in Lou (1998) with the interaction of MHD wind streams,
whilst the knot's luminosity can be maintained by particle injection from the pulsar wind. 
Variations in the knot's luminosity and morphology are anticipated
as successive quasi-periodic disturbances emanate 
from the pulsar. This being so it argues
for a fundamentally dynamical nature to the observed synchrotron knot
which may only be really discernible using future HST or ground-based 
adaptive optics observations.

Finally, the MERLIN data definitively rules out the putative optical counterparts to
PSR B1951+32 suggested by Butler et al. (2002) and Moon et al. (2004),
and provides an unambiguous error box with which to assist future
high time resolution searches.

\section*{Acknowledgments}
John Cunniffe is thanked for his counsel in certain aspects of the manuscript.
The authors gratefully acknowledge the support of Enterprise Ireland under
grant award SC/2001/0322. SB acknowledges support of an EU Marie Curie
Fellowship whilst at the Jodrell Bank Training Site for Radio
Astronomy which is funded by the EC. The authors are very happy to
acknowledge support of PPARC in the UK and the NRAO in the US
for access to MERLIN and the VLA respectively.


\clearpage

\begin{figure} 
\plotone{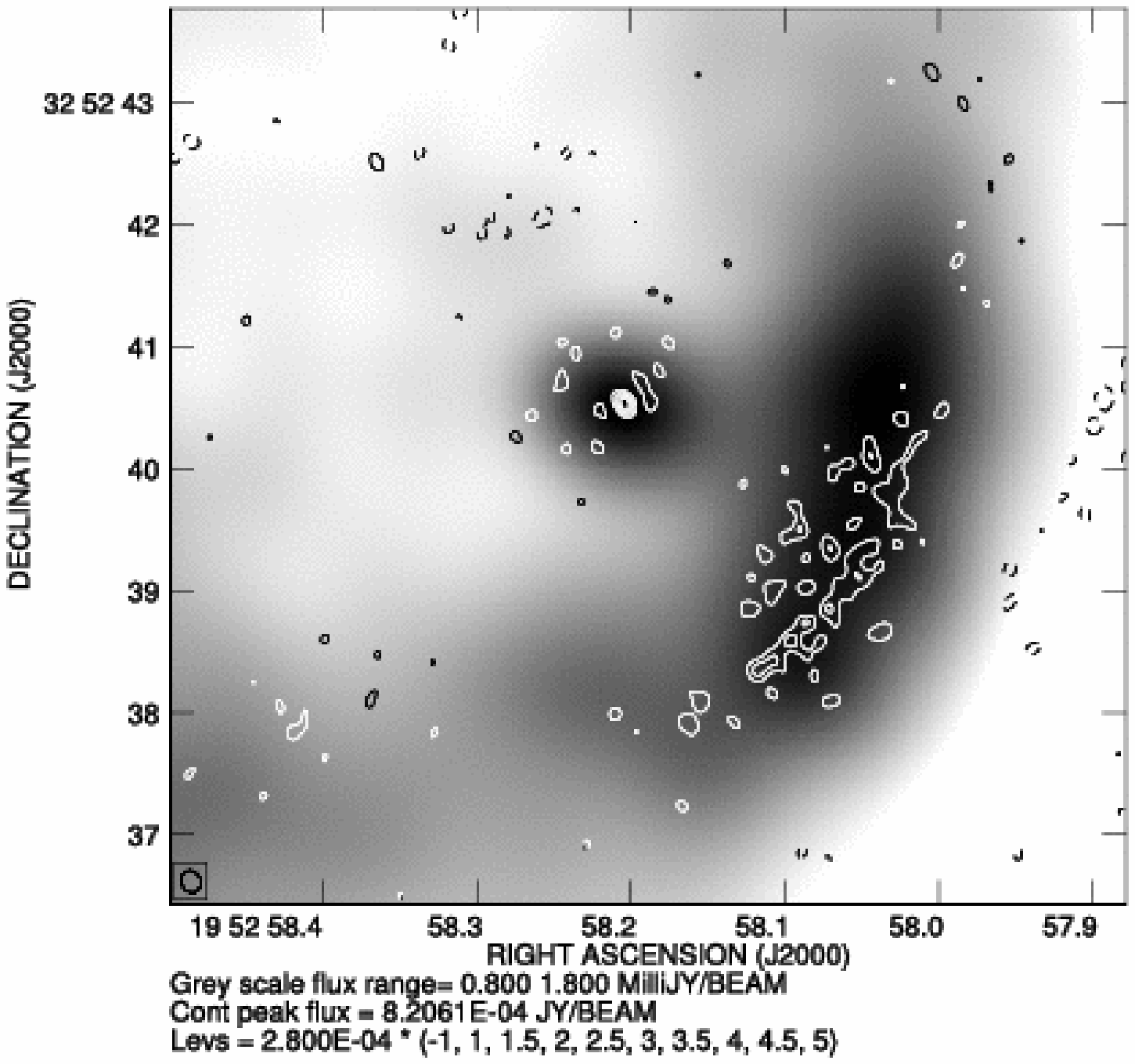}
\vspace{-60mm}
\caption{VLA grey-scale map with superimposed MERLIN L band contour map. The pulsar is
clearly evident in the image centre as both a VLA and MERLIN point
source. The MERLIN resolved contoured shock-structure is to the SW of the pulsar,
coincident with the poorer resolution VLA data, which was obtained 
in July 2003 in the A configuration. The rms noise level for the MERLIN
data is 1.0 $\times$ 10$^{-4}$ Jy/beam, and natural weighting was used in
its analysis.}
\label{merlin2} 
\end{figure}

\clearpage

\begin{figure} 
\plotone{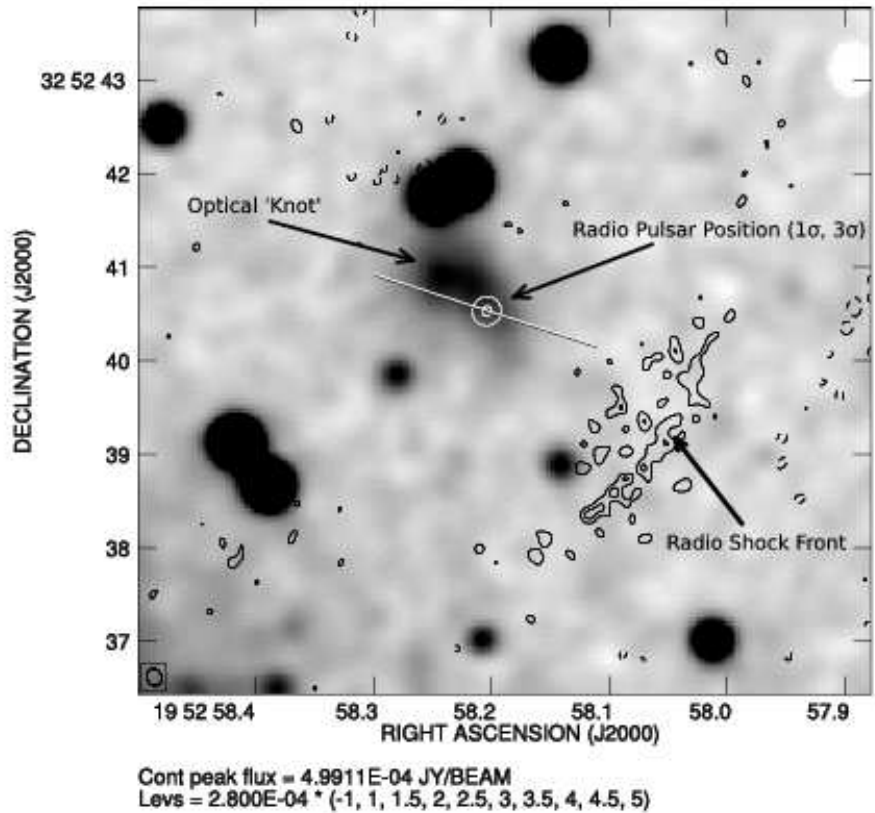}
\vspace{-60mm}
\caption{Smoothed inverse grey-scale HST F547M image of the central region, 
showing the `teardrop'
like optical synchrotron knot as originally marked out by Hester (2000).
The white ellipses in the centre show the 1 and 3 $\sigma$
positions of the radio pulsar with respect to the optical frame as defined
by the WCS. The black
contours show the MERLIN map (the pulsar has been removed for clarity.)
The white line shows the proper motion of the pulsar over a 100yr time frame.
(from Migliazzo et al. (2002).)
}
\label{hst1} 
\end{figure}

\end{document}